\documentclass[12pt]{article}

\def\be{\begin{eqnarray}}
\def\ee{\end{eqnarray}}
\def\eps{\varepsilon}

\begin{document}
\title{Fractional statistics and finite bosonic system: A one-dimensional case}
\author{Andrij Rovenchak\footnote{E-mail: {\tt
andrij@ktf.franko.lviv.ua}}\\
Department for Theoretical Physics, \\
Ivan Franko National University of Lviv,\\
12 Drahomanov St., Lviv, UA-79005, Ukraine}

\maketitle

\begin{abstract}
The equivalence is established between the one-dimensional (1D)
Bose-system with a finite number of particles and the system obeying the fractional
(intermediate) Gentile statistics, in which the maximum occupation of single-particle
energy levels is limited. The system of 1D harmonic oscillators is
considered providing the model of harmonically trapped Bose-gas.
The results are generalized for the system with power energy spectrum.


PACS numbers
05.30.Ch, 
05.30.Jp, 
05.30.Pr 

\end{abstract}

\section{Introduction}
The observation of Bose--Einstein condensation (BEC) in ultracold
trapped alkali gases \cite{BEC1,BEC2} gave new stimulus to the study of
this quantum phenomenon. In particular, the effects of a finite number
of particles on BEC of an ideal gas was discussed in theoretical
works \cite{Ketterle96,Deng97,Napolitano97,Pathria98,Li99,Vakarchuk01},
where the corrections to the bulk properties were found.

The aim of the present study is to propose the description of a
bosonic system with a finite number of particles by means of
finding such a model system, for which the treatment might be
mathematically simpler.
To some extent, such an approach has common features with finding
boson--fermion equivalence in ideal gases \cite{Patton05} or
Tonks--Girardeau gas \cite{Girardeau60}
achieved experimentally in 2004 \cite{Peredes04,Kinoshita04}.
The anyon--fermion mapping is also known in the application to ultracold gases
\cite{Girardeau06}.

Haldane's exclusion statistics \cite{Haldane91} was considered by Berg\`ere
\cite{Bergere00}.
The connection of the exclusion (anyon) statistics parameter and the
interaction in one-dimensional systems was studied in
\cite{Isakov94,Murthy94,Ha95}. recently, a combinatorial interpretation of exclusion statistics
was given by Comtet {\it et al.} \cite{Comtet07}.

Another approach is seen in a different type of the fractional statistics,
which is formally understood as an
intermediate one between Fermi and Bose statistics. Namely, the
maximum occupation of a particular energy level is limited to $M$,
with $M = 1$ corresponding to the fermionic distribution and $M =
\infty$ being the bosonic one, respectively. This statistics is
known as the Gentile statistics \cite{Gentile40,Isihara71,Khare97}.
If the relation between the number of particles $N$ in the real system and the
parameter $M$ in the model one can be found, the stated problem is
solved.

For simplicity, a one-dimensional system is considered.
The paper is organized as follows. Microcanonical approach
for harmonic oscillators with single-particle energy
levels given by $\eps_m = \hbar\omega m$ is considered in Sec.~\ref{SecMicro}.
The oscillators, unlike classical particles, are
{\em indistinguishable} reproducing thus a quantum case.
Physically, this corresponds to bosons trapped in a highly
asymmetric harmonic trap.
In Sec.~\ref{SecCanon}, the same system is treated within canonical and grand-canonical
approaches. Sec.~\ref{SecPower} contains the generalization of obtained results
for the system with power energy spectrum $\eps_m \propto m^s$.
Short discussion in Sec.~\ref{SecDiscu} concludes the paper.

\section{Microcanonical approach}\label{SecMicro}

The number of microstates ${\it\Gamma}(E)$ in the system of 1D
oscillators is the number of ways to distribute the energy
$E=\hbar\omega n$ over the (indistinguishable) particles. Such a
problem reduces to the problem in number theory known as the
partition of an integer \cite{Andrews76,Grossman97,Tran04}. An
asymptotic expression for (unrestricted) partition is given by the
well-known Hardy--Ramanujan formula \cite{Hardy18}:
\be\label{Gamma} p(n) = {1 \over 4\sqrt3\,
n}\,e^{\pi\sqrt{2/3}\,\sqrt n}. \ee Thus one obtains:
\be
{\it\Gamma}(E) = {1 \over 4\sqrt3\,
{E/\hbar\omega}}\,e^{\pi\sqrt{2/3}\,\sqrt{E/\hbar\omega}}. \ee
Using the entropy $S=\ln{\it\Gamma}$ from the definition of the
temperature $\displaystyle {1\over T} = {dS\over dE}$ the
following equation of state is obtained: \be\label{ET}
E={\pi^2\over6}\hbar\omega\left(T\over\hbar\omega\right)^2. \ee As
the energy $E$ is an extensive quantity, $E\propto N$, where $N$
is the number of particles, the thermodynamic limit $\omega N=\rm
const$ follows immediately from the above equation. The same
result also might be obtained from different considerations
\cite{Dalfovo99}.

If one considers a finite system of bosons or a system of particles obeying
fractional statistics the number of ways to distribute the energy $E=\hbar\omega n$ over
$N$ particles is the problem of restricted partitions of an
integer number $n$ \cite{Tran04}. For convenience, hereafter $\hbar\omega$ is the unit
of both energy and temperature.

The expression for the finite system is given by the number of
partitions of $n$ into at most $N$ summands and asymptotically
equals \cite{Erdos41}:
\be\label{GammaFin}
{\it\Gamma}_{\rm fin}(n) = {1 \over 4\sqrt3\, n}\,e^{\pi\sqrt{2/3}\,\sqrt n}
\exp\left\{-{\sqrt6\over\pi}\,\sqrt{n}\,e^{-{\pi\over\sqrt6}\,{N\over\sqrt n}}\right\}.
\ee
The result reducing to the fractional statistics
was considered by Srivatsan {\em et al.}\ \cite{Srivatsan06}, it
corresponds to the number of partitions of $n$ where every summand
appears at most $M$ times:
\be\label{GammaFrac}
{\it\Gamma}_{\rm frac}(n) = {1 \over 4\sqrt3\, n}\,e^{\pi\sqrt{2/3}\,\sqrt
n\left(1-{1\over\sqrt M}\right)^{1/2}} \left(1-{1\over\sqrt
M}\right)^{1/2}.
\ee

Comparing the entropies $S_{\rm fin} = \ln{\it\Gamma}_{\rm fin}$ and $S_{\rm frac} = \ln{\it\Gamma}_{\rm frac}$,
one finds the equivalence condition linking the maximum occupation parameter $M$ and
the number of particles $N$:
\be\label{MicroResult}
M\sim\exp\left({\pi\over\sqrt6}\,{N\over\sqrt n}\right).
\ee

\section{Canonical and grand-canonical approach}\label{SecCanon}

It is straightforward to show that in the case of the defined fractional statistics
the occupation number of the energy level $\eps$ equals \cite{Gentile40,Isihara71,Khare97}
\be\label{DistrFun}
f_M(\eps,\mu,T) = {1\over e^{(\eps-\mu)/T}-1} - {M+1\over e^{(M+1)(\eps-\mu)/T}-1},
\ee
where $\mu$ is the chemical potential and $T$ is the temperature.

The chemical potential is related to the number of particles $\cal N$
as follows:
\be\label{Ndef}
{\cal N}=\sum_{j=0}^\infty f_M(\eps_j,\mu,T)
\ee
and energy $E$ equals
\be\label{Edef}
E=\sum_{j=0}^\infty \eps_j f_M(\eps_j,\mu,T).
\ee

However, the case of a finite system is much easier to implement
in the canonical approach. It is possible to show that the
partition function of $N$ indistinguishable 1D oscillators is
given by (cf.~\cite{Grossman97}):
\be
Z_N=\prod_{j=1}^N\left(1-e^{j/T}\right)^{-1},
\ee
from which the energy $E_N$ can be calculated. In the limit of
large $N$ the leading term is given by
\be\label{ENcorr}
E_N-E_{\rm Bose} \sim {Ne^{-N/T}\over e^{1/T}-1},
\ee
where $E_{\rm Bose}$ is the energy of an infinite bosonic system.

For the fractional-statistics system the grand-canonical approach
is used. The fugacity $z=e^{\mu/T}$ is represented as
$z=z_{\rm Bose}+\Delta z$ with $z_{\rm Bose}$ satisfying
\be
{\cal N}=\sum_i {1\over z_{\rm Bose}^{-1}e^{\varepsilon_i/T}-1}.
\ee
It is found that $\displaystyle\Delta z\sim {1\over M}$ in the limit of large $M$,
from which the correction to the energy given by Eq.~(\ref{Edef}) follows:
\be\label{EMcorr}
E_M - E_{\rm Bose} \sim {1\over M}.
\ee
Comparing Eqs.~(\ref{ENcorr}) and (\ref{EMcorr}) one obtains the following relation between
the parameters $M$ and $N$:
\be\label{GrandResult}
M\sim {1\over N}e^{N/T}.
\ee
In the exponent, the temperature $T$ is related to
the energy level $n$ of (\ref{MicroResult}) via Eq.~(\ref{ET}) (with $E=n$).
Result (\ref{GrandResult}) thus reproduces the microcanonical one (\ref{MicroResult})
up to the negligible factor of $1/N$ --- it must be taken into
account that only leading terms were preserved in the logarithms of (\ref{GammaFin}) and
(\ref{GammaFrac}).

\section{Power energy spectrum}\label{SecPower}

In this section, a general power energy spectrum $\eps= a m^s$ ($s>0$) is considered.
By choosing appropriate energy units, one can set the constant $a=1$.
In fact, only $s=1$ and $s=2$ cases are realized in real physical systems \cite{Srivatsan06},
but other values can effectively occur in some exotic model systems.

To obtain ${\it\Gamma}_{\rm fin}(n)$ for arbitrary $s$ it is worth
to recall briefly the derivation of the expression for restricted partitions
from \cite{Tran04}.

Partition function $Z(\beta)$ and the number of microstates ${\it\Gamma}(E)$
are related via the Laplace transform:

\be \label{ZLaplace}
Z(\beta) = \int_0^\infty {\it\Gamma}(E)\, e^{-\beta E}\, dE,
\qquad
{\it\Gamma}(E) = {1\over2\pi i}
\int_{-i\infty}^{+i\infty} Z(\beta)\, e^{\beta E}\, d\beta.
\ee

The entropy $S(\beta)$ equals
\be
S(\beta) = \beta E + \ln Z(\beta).
\ee

For energy spectrum $\eps_m=m^s$ the partition function is
\be
Z(\beta) = \prod_{m=1}^\infty \left(1-e^{-\beta m^s}\right)^{-1}.
\ee

Using the saddle-point method, one can evaluate ${\it\Gamma}(E)$
(\ref{ZLaplace}) as follows:
\be
{\it\Gamma}(E) = {\exp[S(\beta_0)] \over \sqrt{2\pi
S''(\beta_0)}}.
\ee

The entropy $S(\beta)$, after applying the Euler--Maclaurin summation formula, can
be expressed in such a form
\be
S(\beta) = \beta E -\sum_{m=1}^\infty \ln\left(1-e^{-\beta m^s}\right) 
  = \beta E + {C(s) \over \beta^{1/s}} +{1\over2}\ln\beta +\ldots,
\ee
where
\be
C(s) = \Gamma\left(1+{1\over s}\right) \zeta\left(1+{1\over s}\right),
\ee
$\Gamma(z)$ and $\zeta(z)$ being Euler's gamma-function and Riemann's zeta-function,
respectively.

The stationary point $\beta_0$ is
\be
\beta_0 = \left(C(s) \over s E\right)^{s/(s+1)} =
\lambda_s E^{-s/(s+1)}.
\ee
Thus, the number of microstates is
\be
{\it\Gamma}(E) = {\lambda_s \over
(2\pi)^{(s+1)/2}}\sqrt{s\over s+1}E^{-{3s+1\over 2(s+1)}}
\exp\left[\lambda_s(s+1)E^{1\over s+1}\right].
\ee

Substituting $E$ with $n$ one can obtain the well-known Hardy--Ramanujan
formula \cite{Hardy18} for the number of partitions of an integer $n$ into the
sum of $s$th powers.

When the number of particles $N$ in the system is finite,
the correction to the above formula must be found. In this case, the partition function
equals
\be
\ln Z_N(\beta) = -\sum_{m=1}^N \ln\left(1-e^{-\beta m^s}\right)
\ee
and for the entropy one has
\be
S_{\rm fin}(\beta) = \beta E -\sum_{m=1}^N \ln\left(1-e^{-\beta m^s}\right).
\ee
After simple transformations it is easy to obtain the following:
\be
S_{\rm fin}(\beta)
=
S(\beta) - {1\over s\beta^{1/s}} \Gamma\left({1\over s},\beta N^s\right),
\ee
where $\Gamma(a,x)$ is incomplete $\Gamma$-function.
Thus,
\be
{\it\Gamma}_{\rm fin}(E) = {\it\Gamma}(E) \exp\left[-{1\over s\beta_0^{1/s}}
                  \Gamma\left({1\over s},\beta_0
                  N^s\right)\right].
\ee
Applying the asymptotic expansion for $\Gamma(a,x)$ \cite[Eq.~6.5.32]{AbramowitzStegun},
we finally arrive at the following:
\be
{\it\Gamma}_{\rm fin}(E) = {\it\Gamma}(E) \exp\left[-{1\over
s\beta_0} N^{1-s}e^{-\beta_0 N^s}\right].
\ee

Substituting $E$ with $n$ one obtains the result for restricted partitions
\be
{\it\Gamma}_{\rm fin}(n)= {\it\Gamma}(n) \exp\left[- {1\over \lambda_s s}n^{s/(s+1)} N^{1-s}
e^{-\lambda_s N^s n^{-s/(s+1)}}\right],
\ee
cf. also Eq.~(17) from \cite{Comtet07a}.
For this function, the notation $p^s_N(n)$ is traditionally used,
note, however, that in the problem of integer partitions $s$ must
be integer.
For $s=1$ the obtained expression reduces to that of Erd\H{o}s and Lehner \cite{Erdos41},
see Eq.~(\ref{GammaFin}).

The fractional-statistics result can be directly taken from \cite{Srivatsan06}:
\be
{\it\Gamma}_{\rm frac}(n) &=&
{\lambda_s \over
(2\pi)^{(s+1)/2}}\sqrt{s\over s+1} \left(1-{1\over(M+1)^{1/s}}\right)^{s/(s+1)} n^{-{3s+1\over 2(s+1)}}
\nonumber\\
&&{}\times
\exp\left[\lambda_s \left(1-{1\over(M+1)^{1/s}}\right)^{s/(s+1)}(s+1)n^{1\over s+1}\right]
\ee

To obtain the relation between the parameters $M$ and $N$, one can
again consider the entropies $S_{\rm frac} = \ln {\it\Gamma}_{\rm frac}$ and
$S_{\rm fin} = \ln {\it\Gamma}_{\rm fin}$:

\be
S_{\rm frac} - S = -{1\over s\beta_0}N^{1-s}e^{-\beta_0N^s},\qquad
S_{\rm fin} - S = -{s\lambda_s\over 1+s} M^{-1/s},
\ee
where $S=\ln\it\Gamma$.

Dropping the constants, the following result is obtained:
\be\label{Ms}
M^{1/s}\sim n^{1-s\over1+s}N^{s-1}\exp\left\{\lambda_s
n^{-{s\over1+s}}N^s\right\}.
\ee

It is interesting to find in this general case the connection between energy $E$ and
temperature $T$ from the definition
$\displaystyle {1\over T}={dS\over dE}$:
\be
{1\over T} = \lambda_s E^{-{s\over 1+s}} \quad\Rightarrow\quad
E^{s\over1+s}=\lambda_s T.
\ee
Thus, the leading contribution in the relation of $M$ and $N$
(\ref{Ms}) is
\be
M\sim\exp\left({sN^s\over T}\right),
\ee
which is compatible with (\ref{GrandResult}).

\section{Discussion}\label{SecDiscu}
To summarize, the equivalence is established between the finite bosonic system
and the system obeying fractional (intermediate) Gentile statistics in the case
of one-dimensional harmonic trap. This approach is extended to a
general power energy spectrum.
While the expressions for two-dimensional (2D)
partitions are also known \cite{Andrews76}, the application to asymmetric (elliptical) traps
as well as the generalization for arbitrary 2D systems needs additional study.

Interacting systems are of special interest now. Weak interactions
are known not to change the properties of a Bose-system drastically.
Thus, one can use, e.~g., a slightly modified excitation spectrum \cite{Rovenchak07}
and, upon calculating the properties of a model fractional-statistics
system, obtain the results for a finite one from the established equivalence.

\section*{Acknowledgements}
I am grateful to my colleagues Prof.~Volodymyr Tkachuk,
Dr.~Taras Fityo, and Yuri Krynytskyi for useful discussions and comments
regarding the presented material.


\begin{thebibliography}{00}
\bibitem{BEC1}M. H. Anderson {\em et al.},
     Science {\bf 269}, 198 (1995).
\bibitem{BEC2}K. B. Davis {\em et al.},
     Phys. Rev. Lett. {\bf 75}, 3969 (1995).

\bibitem{Ketterle96}W. Ketterle and N. J. van Druten,
     Phys. Rev. A {\bf 54}, 656 (1996).
\bibitem{Deng97}W. Deng and P. M. Hui,
     Solid Stat. Commun. {\bf 104}, 729 (1997).
\bibitem{Napolitano97}R. Napolitano, J. DeLuca, V. S. Bagnato, G. C. Marques,
     Phys. Rev. A {\bf 55}, 3954 (1997).
\bibitem{Pathria98}R. K. Pathria,
     Phys. Rev. A {\bf 58}, 1490 (1998).
\bibitem{Li99}M. Li, L. Chen, J. Chen, Z. Yan, C. Chen,
     Phys. Rev. A {\bf 60}, 4168 (1999).
\bibitem{Vakarchuk01}I. O. Vakarchuk and A. A. Rovenchak,
     Condens. Matter Phys. {\bf 4}, 431 (2001).

\bibitem{Patton05}K. R. Patton, M. R. Geller, and M. P. Blencowe,
     Physica A {\bf 357}, 427 (2005).

\bibitem{Girardeau60}M. Girardeau,
     J. Math. Phys. {\bf 1}, 516 (1960).
\bibitem{Peredes04}B. Peredes {\em et al.},
     Nature {\bf 429}, 277 (2004).
\bibitem{Kinoshita04}T. Kinoshita, T. Wenger, and D. S. Weiss,
     Science {\bf 305}, 1125 (2004).
\bibitem{Girardeau06}M. D. Girardeau,
     Phys. Rev. Lett. {\bf 97}, 100402 (2006).

\bibitem{Haldane91}F. D. M. Haldane,
     Phys. Rev. Lett. {\bf 67}, 937 (1991).
\bibitem{Bergere00}M. C. Berg\`ere,
     J. Math. Phys. {\bf 41}, 7252 (2000).
\bibitem{Isakov94}S. B. Isakov,
     Phys. Rev. Lett. {\bf 73}, 2150 (1994).
\bibitem{Murthy94}M. V. N. Murthy and R. Shankar,
     Phys. Rev. Lett. {\bf 73}, 3331 (1994).
\bibitem{Ha95}Z. N. C. Ha,
     Nucl. Phys. B {\bf 435},604 (1995).
\bibitem{Comtet07}A. Comtet, S. N. Majumdar, and S. Ouvry,
     J. Phys. A. {\bf 40}, 11255 (2007).

\bibitem{Gentile40}G. Gentile, Nuovo Cim. {\bf 17}, 493 (1940).
\bibitem{Isihara71}A. Isihara,
     {\em Statistical Physics} (Academic Press, New-York--London, 1971).
\bibitem{Khare97}A. Khare,
     {\em Fractional Statistics and Quantum Theory}
     (World Scientific, Singapore, 1997).

\bibitem{Andrews76}G. E. Andrews, {\em The Theory of Partitions} (Addison-Wesley, Reading, Mass., 1976).
\bibitem{Grossman97}S. Grossmann and M. Holthaus,
     Phys. Rev. Lett. {\bf 79}, 3557 (1997).
\bibitem{Tran04}M. N. Tran, M. V. N. Murthy, and R. J. Bhaduri,
     Ann. Phys. {\bf 311}, 204 (2004).

\bibitem{Hardy18}G. H. Hardy and S. Ramanujan,
     Proc. London Math. Soc. {\bf 17}, 75 (1918).

\bibitem{Dalfovo99}F. Dalfovo {\em et al.}, Rev. Mod. Phys. {\bf 71}, 463 (1999).

\bibitem{Erdos41}P. Erd\H{o}s and J. Lehner,
     Duke Math. J. {\bf 8}, 345 (1941).

\bibitem{Srivatsan06}C. S. Srivatsan, M. V. N. Murthy and R. K. Bhaduri,
     Pramana -- J. Phys. {\bf 66}, 485 (2006).

\bibitem{AbramowitzStegun}M. Abramowitz, I. A. Stegun,
     {\em Handbook of mathematical functions}. Tenth printing (National Bureau of Standards, 1972).

\bibitem{Comtet07a}A. Comtet, P. Leboeuf, and S. N. Majumdar,
     Phys. Rev. Lett. {\bf 98}, 070404 (2007).

\bibitem{Rovenchak07}A. A. Rovenchak,
     J. Low Temp. Phys. {\bf 148}, 411 (2007).

\end{thebibliography}
\end{document}